\documentclass[12pt]{article}
\usepackage{amssymb}
\usepackage{amsmath}
\usepackage{IEEEtrantools}
\usepackage{pdflscape}
\usepackage{afterpage}
\usepackage{capt-of}

\usepackage{graphicx}
\usepackage{epstopdf}
\usepackage{subfig}

\usepackage{amsfonts,amsthm}
\usepackage{amsmath}  
\usepackage{latexsym}
\usepackage{amssymb}
\usepackage{mathrsfs}
\usepackage{tikz}
\usepackage[all]{xy}
\usepackage[normalem]{ulem}
\usepackage{hyperref}

\usepackage{changepage}

\numberwithin{equation}{section}

\newcommand{\be}{\begin{equation}}
\newcommand{\ee}{\end{equation}}
\newcommand{\bea}{\begin{eqnarray}}
\newcommand{\eea}{\end{eqnarray}}
\newcommand{\non}{\nonumber}

\newcommand{\tr}{\mathop{\rm tr}\nolimits}

\begin{document}

\begin{titlepage}
\strut\hfill UMTG--300
\vspace{.5in}

\begin{center}

\LARGE Free-Fermion entanglement and orthogonal polynomials\\
\vspace{1in}
\large 
Nicolas Cramp\'e \footnote{Institut Denis-Poisson CNRS/UMR 7013 - Universit\'e de Tours - Universit\'e d'Orl\'eans, 
Parc de Grammont, 37200 Tours, France.\ \  crampe1977@gmail.com}${}^{,3}$, 
Rafael I. Nepomechie \footnote{Physics Department,
P.O. Box 248046, University of Miami, Coral Gables, FL 33124 USA.\\ nepomechie@miami.edu} and
Luc Vinet \footnote{Centre de Recherches Math\'ematiques, Universit\'e de Montr\'eal,
P.O. Box 6128, Centre-ville Station,
Montr\'eal (Qu\'ebec), H3C 3J7, Canada.\\vinet@crm.umontreal.ca}  
\\[0.8in]
\end{center}

\vspace{.5in}

\begin{abstract}
We present a simple construction for a tridiagonal
matrix $T$ that commutes with the hopping matrix for the entanglement
Hamiltonian ${\cal H}$ of open finite free-Fermion chains associated
with families of discrete orthogonal polynomials.  It is based on the
notion of algebraic Heun operator attached to bispectral problems, and
the parallel between entanglement studies and the theory of time and
band limiting.  As examples, we consider Fermionic chains related to
the Chebychev, Krawtchouk and dual Hahn polynomials.  For the former
case, which corresponds to a homogeneous chain, the outcome of our
construction coincides with a recent result of Eisler and Peschel; the
latter cases yield commuting operators for particular inhomogeneous
chains.  Since $T$ is tridiagonal and non-degenerate, it can be
readily diagonalized numerically, which in turn can be used to
calculate the spectrum of ${\cal H}$, and therefore the entanglement
entropy.
\end{abstract}

\end{titlepage}

\setcounter{footnote}{0}

\section{Introduction}\label{sec:intro}

Entanglement, a distinctive feature of the quantum realm often
quantified through entropies, is of fundamental relevance in black
hole physics, information theory and many-body problems \cite{Calabrese:2005zw,Calabrese2009,Casini:2009sr,Laflorencie:2015eck}.  It is hence
actively studied in a variety of situations.  This paper relates to
entanglement in free-Fermion or solvable XX spin
chains, a topic that is generating much attention on its own (for a
review see for instance \cite{2009JPhA...42X4002L}).  Basically, the
question is the following: Suppose the whole chain is in the quantum
(pure) state described by $|\Psi\rangle\!\rangle$, which we shall here take to
be the ground state. We divide the chain 
in two spatial parts $1$ and $2$, and ask
how are these parts coupled in $|\Psi\rangle\!\rangle$.  Since all the
properties of the subsystem $1$ are provided by the reduced density matrix
$\rho_{1}$ obtained by tracing $|\Psi \rangle\!\rangle 
\langle\!\langle \Psi|$ over part 2, therein will be all the entanglement
information.  For example, the von 
Neumann entropy is given by $S_{1} = -\tr
(\rho_{1} \log \rho_{1})$, which amounts to finding the eigenvalues of
$\rho_{1}$.

In the following we shall take our subsystem $1$ to consist of the first
consecutive $l+1$ sites of the chain labelled by $n = 0, 1, ...  , N$.
Diagonalizing the $2^{l+1} \times 2^{l+1}$ reduced density matrix 
$\rho_{1}$ could become
prohibitive as $l$ grows.  Fortunately, owing to the fact that the
eigenstates of the chains considered are Slater determinants, it has
been shown \cite{2003PhRvL..90v7902V, 2003JPhA...36L.205P} that
$\rho_{1}$ 
in a chosen state can be obtained
from the 1-particle correlation matrix $C$ in that state, thus reducing
the determination of the entanglement entropy to finding the
eigenvalues of a $(l+1) \times (l+1)$ matrix.  Furthermore, it was
observed \cite{2003JPhA...36L.205P, 2009JPhA...42X4003P} that
as a consequence, $\rho_{1}$ must be of the thermodynamic form
\begin{equation}
\rho_{1} = \kappa \; \exp (-\mathcal{H}) \,,
\label{entH}
\end{equation}
where $\mathcal{H}$, known as the \textit{Entanglement Hamiltonian},
is also Fermionic (but is not the Hamiltonian of the subsystem).  The
constant $\kappa$ simply ensures normalization, i.e. $\tr \rho_{1} = 1$.
The hopping matrix $h$ that characterizes $\mathcal{H}$ \footnote{We 
shall reserve the term ``hopping 
matrix'' for the coefficients appearing in the Entanglement Hamiltonian, 
rather than in the original Hamiltonian, see \eqref{hopping} below.}
is hence a function of the correlation matrix $C$.  It remains, however, that the
eigenvalue problem for $h$ or equivalently $C$ becomes numerically
quite difficult as $l$ grows, because these are full matrices with
closely spaced eigenvalues.  As pointed out in \cite
{2006PhRvL..96j0503G} and stressed by Peschel and Eisler \cite
{2004JSMTE..06..004P, 2013JSMTE..04..028E, 2018arXiv180500078E}, 
classical results in signal processing (as
well as in random matrix theory) can be brought to bear on the
analysis of the entanglement properties of free-Fermion chains in
certain instances.  Since this is directly related to the main results
that are reported in this paper, let us briefly offer here some
relevant background.

In its initial form, the theory of \textit {Time and Band Limiting}
developed by Slepian, Landau and Pollack aims to determine an unknown
function/signal with two kinds of limitations: (i) the duration of the
transmission interval is finite and known, and (ii) only a piece of the
function's Fourier transform, say over a certain band of frequencies,
is available.  The optimal use of this information requires finding
the eigenfunctions of certain integral operators whose non-local
character makes the numerical analysis almost 
intractable.  Amazingly,
Slepian et al.  have circumvented this problem by finding a
differential operator that commutes with the integral one, that thus
shares with it common eigenfunctions, and that has eigenvalues that are
nicely spread.  The original work has been generalized in various
directions and is having numerous applications.  For reviews the
reader could consult \cite{MR710468, Landau1985}.

The reasons and the circumstances for the existence of commuting
operators in time and band limiting problems are still not fully
understood.  This has motivated in part the seminal work of
Duistermaat and Gr\"unbaum on bispectral problems
\cite{1986CMaPh.103..177D}.  With their three-term recurrence relation
and their differential/difference equation, the hypergeometric
polynomials (which are organized in a hierarchical way in the so-called Askey scheme \cite{Koekoek,Koekoek2010}) 
are  prominent examples
of bispectral problems.  Over the years, with his collaborators and
students, Gr\"unbaum has discovered and developed many realizations of
limiting problems with commuting operators.  In \cite{MR716099} for
example, working in the framework of the classical orthogonal
polynomials (Jacobi, Laguerre, Hermite), he has found an analog of the
results obtained by Slepian et al.  with the Fourier transform.  This
has been extended to more general orthogonal polynomials in
\cite{doi:10.1137/0604012, doi:10.1137/0515081}.  The questions
regarding the origin of the commuting operator were recently revisited
in \cite{2018CMaPh.364.1041G}; the concept of algebraic Heun 
operator\footnote{The reason for the name is that, when applied
\cite{2017JMP....58c1703G} to the bispectral operators of the Jacobi
polynomials, the construction precisely yields the Heun equation with
four regular Fuchsian singularities \cite{Kristensson},
\cite{Ronveaux}.} attached to bispectral problems was introduced and
it was shown that, generically, commuting operators of time and band
limiting problems belong to that class of so-called Heun operators
thus rediscovering and extending beyond the finite-dimensional case, a
result of Perline \cite{Perline:1987:DTL:37170.37175}.  It is this
simple construction that we here wish to apply to the search of
tridiagonal matrices that commute with the hopping matrix
for finite free-Fermion chains that are associated to orthogonal
polynomials of the Askey scheme.

The relevance of the time and band limiting theory to the study of the
entanglement properties of free-Fermion chains is now readily seen.
Restricting to a subsystem, i.e. to the first $l+1$ sites of the 
chain, corresponds to limiting time.  Filling the Fermi sea (or exciting a
consecutive set of 1-particle energy eigenvalues) is tantamount to band limiting.
The set-ups are clearly parallel.  The correlation matrix $C$ is the
operator that we wish to diagonalize; and its analysis would much
benefit from knowing a tridiagonal matrix $T$  that commutes
with it, or equivalently with the hopping matrix.  We shall
point out in what follows that the formula of Perline 
\cite{Perline:1987:DTL:37170.37175} which
specializes the corresponding algebraic Heun operator, readily
provides this commuting Jacobi matrix when the chain is associated to
orthogonal polynomials of the Askey scheme and the subsystem
corresponds to the first $l +1$ sites of the chain and the filling is
done with consecutive ``momenta''.  The key point will be to recognize
and exploit the presence in these situations of the second operator of
the bispectral pair.  In a recent study \cite{2018arXiv180500078E}
focused on finite free-Fermionic chains with uniform couplings, Eisler
and Peschel have obtained the tridiagonal matrix that commutes with the
hopping matrix.  
They have found that it coincides with the
results obtained by Gr\"unbaum in \cite{doi:10.1137/0602017} and 
observed that the expression for the commuting
 $T$ corresponds to what is obtained from
conformal field theory \cite{Cardy:2016fqc}. 
In the following we shall indicate how this tridiagonal commuting matrix is
straightforwardly obtained by applying the algebraic Heun construction
to a truncation of the Chebychev polynomials of the second kind.  We
note that there is currently interest also in the study of
inhomogeneous chains from the entanglement viewpoint (see for instance
\cite{2010NJPh...12k3049V, 2018JSMTE..04.3105T}).  The method
highlighted in this paper also lends itself to certain chains of that
type, and again easily provides a tridiagonal matrix 
that commutes with the hopping matrix for the entanglement Hamiltonian.  
This is done by connecting with
hypergeometric orthogonal polynomials, and will be illustrated for two
inhomogeneous free-Fermionic chains respectively associated to the
Krawtchouk and the dual Hahn polynomials.

The remainder of the paper is organized as follows.  In Section 2, we
shall introduce the Hamiltonians of the finite free-Fermionic chains
that will be considered.  How their eigenstates are obtained from the
one-excitation dynamics will be reviewed and the required
diagonalization using orthogonal polynomials will be explained.  The
ground state in which entanglement will be studied, shall be given in
Section 3 where the connections between the 1-particle correlation
matrix, the entanglement Hamiltonian and the reduced density matrix
will be reviewed.  With an eye to considering chains with couplings
given by the recurrence coefficients of various families of discrete
orthogonal polynomials, we shall recall in Section 4 properties that
will be used.  The construction from the algebraic Heun operator of
the operator that commutes with the hopping matrix of the entanglement Hamiltonian will be
described in Section 5, and will be seen to exploit the bispectrality
of the underlying polynomials.  Section 6 will be dedicated to the
finite free-Fermion spin chain with uniform couplings and to
recovering from the algebraic Heun operator approach applied to
truncated Chebychev polynomials, the commuting tridiagonal matrix
obtained in \cite{2018arXiv180500078E, doi:10.1137/0602017}.  
Section 7 will present two inhomogeneous free-Fermionic
chains respectively associated to the Krawtchouk and dual Hahn
polynomials together with the tridiagonal matrices commuting with the
corresponding 
hopping matrices.  Finally, Section 8 will
offer concluding remarks.

\section{Free-Fermion chains and their diagonalization}

We consider the following open quadratic free-Fermion inhomogeneous Hamiltonian 
\begin{eqnarray}\label{eq:Hff}
\widehat{\mathcal{H}}=\sum_{n=0}^{N-1}J_{n} (  c_n^\dagger c_{n+1} + c_{n+1}^{\dagger} c_{n})- \sum_{n=0}^{N}B_n c_{n}^{\dagger} c_{n}=\sum_{m,n=0}^N c^\dagger_m \widehat H_{mn}c_n \ ,
\end{eqnarray}
where $J_n$ and $B_n$ are real parameters, and $\{ c_{m}^{\dagger} 
\,, c_{n} \} = \delta_{m,n}$. 
For the sake of simplicity of the following computations, we enumerate the sites of the lattice from $0$ to $N$.
Let us remark that the Hamiltonian \eqref{eq:Hff} can be obtained 
by a Jordan--Wigner transformation from the following XX model
\begin{equation}
 \widehat{\mathcal{H}} = -\frac{1}{2}\sum_{n=0}^{N-1}J_{n} (\sigma^x_n\sigma^x_{n+1}+ \sigma^y_n\sigma^y_{n+1})-\frac{1}{2} \sum_{n=0}^{N}B_n \sigma^z_{n}\ ,
\end{equation}
with $c_n^\dagger=\sigma^z_0\dots \sigma_{n-1}^z 
\sigma_n^+$ and $c_n=\sigma^z_0\dots \sigma_{n-1}^z \sigma_n^-$.

In order to diagonalize $\widehat{\mathcal{H}}$, 
it is convenient to first diagonalize the $(N+1)\times (N+1)$ matrix $\widehat H= | \widehat H_{mn}|_{0\leq m,n \leq N}$.
In the canonical orthonormal basis $\{|0\rangle,|1\rangle,\dots ,|N\rangle \} $ 
of $\mathbb{C}^{N+1}$, called the position basis, $\widehat H$ acts as follows
\begin{equation}\label{eq:Hh}
 \widehat H |n\rangle =J_{n-1} |n-1\rangle  -B_n |n\rangle  + J_{n} 
 |n+1\rangle\,, \qquad 0\leq n \leq N\,, 
\end{equation}
with the convention $J_{N}=J_{-1}=0$. It takes the tridiagonal form
\begin{eqnarray}
 \widehat H=\begin{pmatrix}
             -B_0 & J_0 & \\
             J_0 & -B_1 & J_1 \\
             & J_1 & -B_2 & J_2 \\
             && \ddots & \ddots & \ddots \\
            &&&J_{N-2} & -B_{N-1} & J_{N-1}\\
             &&&& J_{N-1} & -B_N
            \end{pmatrix}\ .
\end{eqnarray}
The spectral problem for $\widehat H$ reads
\begin{equation}
 \widehat H |\omega_k\rangle = \omega_k |\omega_k\rangle\ ,
\end{equation}
where
\begin{equation}
 |\omega_k\rangle =\sum_{n=0}^N \phi_n(\omega_k) |n\rangle\ .
\end{equation}
We choose to order the $N+1$ eigenvalues $\omega_0, \omega_1, \dots 
\omega_N$ such that $\omega_k < \omega_{k+1}$.
We also choose $|\omega_0\rangle,|\omega_1\rangle, \dots 
|\omega_N\rangle$ such that they form an orthonormal basis of 
$\mathbb{C}^{N+1}$, called the momentum basis. The eigenfunctions 
$\phi_n(\omega_k)$ are real, since the matrix $\widehat H$ is 
real and its eigenvalues are non-degenerate (see e.g. Lemma 3.1 in 
\cite{2004math......8390T}, we suppose that $J_n\neq 0$).
Therefore, the eigenfunctions satisfy the orthonormality conditions
\begin{equation}
 \sum_{n=0}^N \phi_n(\omega_k) \phi_n(\omega_p)=\delta_{kp}\,.
 \label{orthophi}
\end{equation}
From relation \eqref{eq:Hh}, we deduce that $\phi_n(\omega_k)$ must satisfy the following recurrence relation
\begin{equation}
\omega_k \phi_n(\omega_k) = J_{n}\phi_{n+1}(\omega_k) - B_n \phi_n(\omega_k) 
+  J_{n-1} \phi_{n-1}(\omega_k)\,, \qquad 
 0\leq n\leq N \,.
 \label{recurphi}
\end{equation}

Having diagonalized $\widehat H$, we see that the Hamiltonian 
$\widehat{\mathcal{H}}$ \eqref{eq:Hff} can be rewritten as 
\begin{equation}
\widehat{\mathcal{H}}=\sum_{k=0}^{N} \omega_{k} 
\tilde{c}^{\dagger}_{k} \tilde{c}_{k} \,, 
\end{equation}
where the annihilation operators $\tilde{c}_{k}$ are defined by
\begin{equation}
\tilde{c}_{k} = \sum_{n=0}^{N} \phi_{n}(\omega_{k})\, 
c_{n} \,, \qquad
c_{n} = \sum_{k=0}^{N} \phi_{n}(\omega_{k})\, 
\tilde{c}_{k} \,,
\label{ctilde}
\end{equation}
and the corresponding relations for the creation operators  
$\tilde{c}^{\dagger}_{k}$ are given by 
Hermitian conjugation of \eqref{ctilde}.
These operators obey the anticommutation relations
\begin{equation}
	\{ \tilde{c}^{\dagger}_{k} \,, \tilde{c}_{p} \} = \delta_{k,p} 
	\,, \qquad  
\{ \tilde{c}^{\dagger}_{k} \,, \tilde{c}^{\dagger}_{p} \} = \{ \tilde{c}_{k} \,, 
\tilde{c}_{p} \} = 0 \,.
\label{CR}
\end{equation}
The eigenvectors of $\widehat{\mathcal{H}}$ are therefore given by
\begin{equation}
	|\Psi\rangle\!\rangle = \tilde{c}_{k_{1}}^\dagger \ldots \tilde{c}_{k_{r}}^\dagger |0\rangle\!\rangle \,,
	\label{grounstate}
\end{equation}
where $k_{1}, \ldots, k_{r} \in \{0, \ldots, N\}$ are pairwise distinct, and
the vacuum state $|0\rangle\!\rangle$ is annihilated by all the annihilation 
operators
\begin{equation}
	\tilde{c}_{k}|0\rangle\!\rangle = 0\,, \qquad k = 0\,, \ldots\,, 
	N \,.
\label{vacuum}
\end{equation}
	The corresponding energy eigenvalues are simply given by
\begin{equation}
	E = \sum_{i=1}^{r} \omega_{k_{i}} \,.
\end{equation}

\section{The entanglement Hamiltonian}

For the sake of concreteness, we shall consider entanglement in the
ground state, which is described below.  We shall further review how
the reduced density matrix for the first $l+1$ sites of the
chain is determined by the 1-particle correlation matrix, and its
relation to the entanglement Hamiltonian.  The parallel with the time
and band limiting problem will also be drawn.

\subsection{Defining the ground state or band limiting}\label{subsec:chop}

The fact that the ground state is constructed by filling the Fermi sea leads to a
``chopping'' in frequency. Indeed, the ground state $|\Psi_{0}\rangle\!\rangle$ of 
the Hamiltonian \eqref{eq:Hff} is given by
\begin{equation}
	|\Psi_{0}\rangle\!\rangle = \tilde{c}_0^\dagger \ldots \tilde{c}_K^\dagger |0\rangle\!\rangle \,,
\end{equation}
where $K\in \{0,1,\dots,N\}$ is the greatest integer below the Fermi
momentum, such that
\begin{equation}
	\omega_{K} < 0\,, \qquad \omega_{K+1} > 0 \,.
\end{equation}	
Let us remark that $K$ can be modified by adding a constant
term in the external magnetic field $B_n$.  

The correlation matrix $\widehat C$ in the ground state is an $(N+1)
\times (N+1)$ matrix with the following entries
\begin{equation}
\widehat C_{mn}=\langle\!\langle \Psi_{0}| c_m^\dagger c_n |\Psi_{0} \rangle\!\rangle \,.
\label{Cmatrixdef}
\end{equation}
Expressing everything in terms of annihilation and creation operators
using \eqref{ctilde} and \eqref{grounstate}, and then using the
anticommutation relations \eqref{CR} and the property \eqref{vacuum} 
of the vacuum state, we obtain
\begin{equation}
\widehat C_{mn}=\sum_{k=0}^K 
\phi_m(\omega_k) \phi_n(\omega_k)\,, \qquad 0\leq n,m\leq N \,.
\label{Cmatrix}
\end{equation}
It is then easy to see that 
\begin{equation}
 \widehat C = \sum_{k=0}^K  |\omega_k\rangle \langle \omega_k| \,,
 \label{Cdag}
\end{equation}
namely, that $\widehat C$ is the projector onto the subspace of
$\mathbb{C}^{N+1}$ spanned by the vectors $|\omega_k\rangle$ with
$k=0,...,K$ running over the labels of the excitations in the ground
state.

\subsection{Space limiting and entanglement}

In order to examine entanglement, we must first define a bipartition
of our free-Fermionic chain.  This is the space limiting.  As
subsystem we shall take the first $\ell+1$ consecutive sites, and shall
find how it is intertwined with the rest of the chain in
the ground state $|\Psi_{0}\rangle\!\rangle$.  To that end, we need the
reduced density matrix 
\begin{equation}
\rho_{1} = \tr_{2} |\Psi_{0}\rangle\!\rangle 
\langle\!\langle \Psi_{0}| \,,
\end{equation}
where part $2$, the complement of part $1$, is comprised of the sites $\{ \ell+1, \ell+2,...,N\}$.

It has been observed that this reduced density matrix is determined by
the spatially ``chopped'' correlation matrix $C$ , 
which is the following $(\ell+1) \times (\ell+1)$ submatrix of $\widehat C$: 
\begin{equation}
  C =  |\widehat C_{mn}|_{0\leq m,n \leq \ell}\,. 
\end{equation}
The argument which we take from \cite{2003JPhA...36L.205P} (see also
\cite{2009JPhA...42X4003P}) goes as follows.  Because the ground state
of the Hamiltonian $\widehat{\mathcal{H}}$ is a Slater determinant,
all correlations can be expressed in terms of the one-particle
functions, i.e. in terms of the matrix elements of $\widehat C$.  When
all the sites belong to the subsystem, since 
\begin{equation}
C_{mn} = \tr (\rho_{1} \; c_m^\dagger c_n)\,,  \qquad  m\,, n \in \{0, 
1,\dots , \ell\},
\label{obs}
\end{equation}
the factorization property will hold according to Wick's theorem if 
$\rho_{1}$ is of the form \eqref{entH} with the entanglement Hamiltonian $\mathcal{H}$ given by
\begin{equation}
\mathcal{H} = \sum_{m,n \in \{0, \dots, \ell\}} h_{mn} \, c_{m}^\dagger 
c_{n} \,.
\label{hopping}
\end{equation}
The hopping matrix $h= |h_{mn}|_{0 \le m,n \le \ell}$ is defined so that \eqref{obs} holds, and one 
finds through diagonalization that 
\begin{equation}
h = \log [(1 - C) /C] \,.
\end{equation}
We thus see that $\rho_{1}$, and hence the entanglement Hamiltonian
$\mathcal{H}$, are obtained from  the $(l+1) \times (l+1)$ matrix $C$.

To calculate the entanglement entropies one therefore has to compute the
eigenvalues of $C$.  As explained in \cite {2004JSMTE..06..004P}, this
is not easy to do numerically because the eigenvalues of that matrix
are exponentially close to $0$ and $1$.  This motivates the search for
a tridiagonal matrix $T$ such that
\begin{equation}
 [T,C] =0 \,.
 \label{want}
\end{equation}
The parallel between the study of entanglement properties of finite
free-Fermion chains and finite-dimensional analogs of time and band
limiting problems indicates that this can be achieved.  
Our aim here is to
show that methods developed in the later context can advantageouly be
used in the former framework.

Introducing the projectors 
\begin{equation}
 \pi_1=\sum_{n=0}^\ell  |n\rangle \langle n| \quad \text{and}\quad  
 \pi_2=\sum_{k=0}^K   |\omega_k\rangle \langle \omega_k| = \widehat C \,,
\end{equation}
the chopped correlation matrix can be written as (see for instance  \cite{Lee:2014nra,2012PhRvB..86x5109H})
\begin{equation}
 C =\pi_1 \pi_2 \pi_1 \ .
 \label{Cpi}
\end{equation}
This makes the limiting explicit.
We shall hence find a $T$ satisfying  \eqref{want}
by looking for a tridiagonal matrix commuting with both projectors:
\begin{equation}
[T,\pi_1] = [T,\pi_2]=0 \,.
\label{pi}
\end{equation}
We may observe that the matrix $D$ defined by $D = \pi_2 \pi_1 \pi_2$
would describe a dual entanglement situation where the 
vacuum state \eqref{vacuum}
would be filled with excitations labelled by the set $\{0, \dots
, \ell\}$, and the subsystem would consist of the sites $\{0, \dots , K\}$.
Since $C$ and $D$ have the same non-zero eigenvalues, the entanglement
entropies will be the same in these two instances.  Such dualities
have been studied in \cite{2017NatSR...711206C}.  We remark that the
$T$ commuting with $C$ will also satisfy $[T, D]=0$
because of \eqref{pi} (see also \cite {2013JSMTE..04..028E}).

\section{Bispectral properties of discrete orthogonal polynomials of the Askey scheme}

A family of discrete orthogonal polynomials $\{ R_{n}(\lambda(x)) \}$ 
with $n, x = 0, 1, \ldots, N$\,, is a sequence of polynomials of degree $n$ in the variable $\lambda(x)$, 
that are orthogonal with respect to some discrete measure
\begin{equation}
	\sum_{x=0}^{N} W(x)\, R_{m}(\lambda(x))\, R_{n}(\lambda(x)) = 
	U_{n} \delta_{m 
	n} \,, \qquad W(x) > 0 \,, \quad U_{n} > 0 \,.
\label{orthoW}	
\end{equation}
We assume the normalization $R_{0}(\lambda(x))=1$.
We consider such polynomials that satisfy a recurrence relation of the form
\begin{equation}
\lambda(x) R_{n}(\lambda(x)) = A_{n} R_{n+1}(\lambda(x)) - \left(A_{n} 
+ C_{n}\right) R_{n}(\lambda(x)) + C_{n} R_{n-1}(\lambda(x)) \,, \quad 
0 \le n \le N\,,
\label{recurgen}
\end{equation}
with $C_{0} = A_{N} = 0$; 
as well as a difference relation of the form
\begin{equation}
f(n) R_{n}(\lambda(x)) = \overline{A}(x) R_{n}(\lambda(x+1)) - \left[ \overline{A}(x) + 
\overline{C}(x) \right] R_{n}(\lambda(x)) +  \overline{C}(x) R_{n}(\lambda(x-1)) \,, \quad
0 \le x \le N\,,
\label{diffgen}
\end{equation}
with $\overline{C}(0) = \overline{A}(N) = 0$. 
A useful reference for such polynomials is \cite{Koekoek,Koekoek2010}, which provides standard examples of bispectral problems
where one has functions $\psi(x,n)$ that are eigenfunctions with eigenvalues depending on $x$ of an operator $L$ acting on 
the variable $n$, and are eigenfunctions as well with eigenvalues depending conversely on $n$ of an operator $Z$ acting on the 
variable $x$. This is the central framework that we shall deal with.

Our basic strategy is to engineer the parameters $J_n$ and 
$B_n$ in the Hamiltonian \eqref{eq:Hff} in such a way that the 
recurrence relation \eqref{recurphi} for the eigenfunctions $\phi_n(\omega_k)$ can be 
mapped to the recurrence relation \eqref{recurgen} for some 
discrete orthogonal polynomials $R_{n}(\lambda(x))$. We then exploit 
the corresponding difference relation \eqref{diffgen} to construct the sought-after 
operator $T$ satisfying \eqref{want}, as explained in Sec. 
\ref{section:Heunn} below.

In practice, we typically start from the recurrence relation 
for a given set of discrete orthogonal polynomials from \cite{Koekoek,Koekoek2010}, and use 
it to determine the parameters $J_n$ and $B_n$. To this end, we set 
\begin{equation}
R_{n}(\lambda(x)) = \frac{\alpha_{n}}{\sqrt{W_{k}}}\, \phi_n(\omega_k) \,,
\label{Rphi}
\end{equation}
where $\alpha_{n}$ are still to be determined. While 
$R_{n}(\lambda(x))$ is a polynomial, $\phi_n(\omega_k)$ is generally 
{\em not} a polynomial, as it contains a transcendental factor that 
is proportional to $\sqrt{W_{k}}$.
We observe that the recurrence relations \eqref{recurphi} and \eqref{recurgen}
can be mapped into each other by means of the identifications
\begin{equation}
	J_{n-1} = \frac{\alpha_{n}}{\alpha_{n-1}} A_{n-1} = 
	\frac{\alpha_{n-1}}{\alpha_{n}} C_{n}   \,, \qquad k = x \,, \qquad 
	\omega_k = \lambda(x) \,, \qquad  W_{k} = W(x) \,.
\end{equation}
It follows that 
\begin{equation}
	\alpha_{n} = \alpha_{n-1}\, \varepsilon 
	\sqrt{\frac{C_{n}}{A_{n-1}}} \,,
\end{equation}
where $\varepsilon = \pm 1$. Solving for the $\alpha$'s, we obtain
\begin{equation}
	\alpha_{n} = \alpha_{0}\, \varepsilon^{n} 
	\prod_{k=1}^{n} \sqrt{\frac{C_{k}}{A_{k-1}}} \,.
	\label{alphas}
\end{equation}
In particular, we arrive at the important result 
that the parameters defining the Hamiltonian \eqref{eq:Hff} are given by
\begin{equation}
	J_{n} = \varepsilon \sqrt{A_{n} C_{n+1}} \,, \qquad B_{n} = A_{n} 
	+ C_{n} \,,
	\label{JBparams}
\end{equation}
where $A_{n}$ and $C_{n}$ are the known coefficients in the recurrence 
relation \eqref{recurgen} for a given family of discrete orthogonal 
polynomials.

The difference relation \eqref{diffgen} for $R_{n}(\lambda(x))$
implies that the eigenfunctions $\phi_n(\omega_k)$ obey the 
corresponding equation
\begin{equation}
\lambda_n \phi_n(\omega_{k}) = \overline{J}_{k} \phi_n(\omega_{k+1}) - \overline{B}_{k} \phi_n(\omega_{k}) 
+  \overline{J}_{k-1} \phi_n(\omega_{k-1}) \,, \qquad 0\leq k\leq N \,,
\label{diffphi}
\end{equation}
with $\overline{J}_{-1}=\overline{J}_{N}=0$, where the coefficients 
are given by\footnote{The consistency condition
\begin{equation}
\frac{\overline{A}(k)}{\overline{C}(k+1)}	= \frac{W_{k+1}}{W_{k}} \non
\end{equation}
is a consequence of the fact that the difference operator is symmetrizable.
}
\begin{equation}
\overline{J}_{k} = \overline{A}(k)\sqrt{\frac{W_{k}}{W_{k+1}}} = 
\overline{C}(k+1)\sqrt{\frac{W_{k+1}}{W_{k}}}\,, \qquad 
\overline{B}_{k} = \overline{A}(k) + 
\overline{C}(k)\,, \qquad \lambda_n = f(n) \,.
\end{equation}

\section{Algebraic Heun operator and commuting tridiagonal matrices}\label{section:Heunn}

The fact that the eigenfunctions $\phi_n(\omega_{k}) = \langle 
n|\omega_{k}\rangle$ obey the
difference relation \eqref{diffphi} can now be exploited to define 
an operator $\widehat X$ in the basis $\{|n \rangle \}$ by
\begin{equation}
 \widehat X |n\rangle = \lambda_n |n\rangle\,,
 \label{Xmat}
\end{equation}
which consequently acts as follows in the $\{  |\omega_{k} \rangle  
\}$ basis
\begin{equation}
 \widehat X |\omega_k\rangle =\overline{J}_{k-1} |\omega_{k-1} \rangle  
 - \overline{B}_{k}|\omega_{k}\rangle+ \overline{J}_{k} |\omega_{k+1}\rangle \ .
\end{equation}
The operators $\widehat H$ and $ \widehat X$ thus 
form a Leonard pair
\cite{2004math......8390T}, meaning roughly that for these two operators there
exist two bases such that in one, $\{ |\omega_k \rangle \}$, $\widehat
H$ is diagonal and $ \widehat X$ is tridiagonal and in the other, $\{
|n \rangle \}$, conversely $\widehat H$ is tridiagonal and $ \widehat
X$ is diagonal.

We may now introduce the algebraic Heun operator defined in
\cite{2018CMaPh.364.1041G} as the most general bilinear expression in
the two bispectral operators $\widehat H$ and $ \widehat X$:
\begin{equation}
 \widehat T= \{ \widehat X , \widehat H \}  + \tau [ \widehat X , 
 \widehat H] + \mu \widehat X + \nu  \widehat H \,,
 \label{Heun}
\end{equation}
where as usual $\{ \widehat X , \widehat H \} = \widehat X \widehat H 
+ \widehat H \widehat X$.  At this point the parameters $\tau, \mu, \nu$
are free.  (Note that allowing for redefinition by an irrelevant
overall factor, the coefficient of $\{ \widehat X , \widehat H \}$ has
been set to 1.)
It is immediate to see that $\widehat{T}$ is tridiagonal in both the position basis
\begin{eqnarray}
  \widehat T |n\rangle&=&J_{n-1}(\lambda_{n-1}(1+\tau)+\lambda_n(1-\tau)+\nu)|n-1\rangle +(\mu \lambda_n-2B_n\lambda_n-\nu B_n)|n\rangle\nonumber \\
 &&+J_n(\lambda_n(1-\tau)+\lambda_{n+1}(1+\tau)+\nu)|n+1\rangle  \ ,
 \label{Tpos}
\end{eqnarray}
and the momentum basis
\begin{eqnarray}
 \widehat T |\omega_k \rangle&=&\overline J_{k-1}(\omega_{k-1}(1-\tau)+\omega_k(1+\tau)+\mu)|\omega_{k-1}\rangle +(\nu \omega_k-2 \overline B_k\omega_k-\mu \overline B_k)|\omega_k\rangle\nonumber \\
 &&+\overline J_k(\omega_k(1+\tau)+\omega_{k+1}(1-\tau)+\mu)|\omega_{k+1}\rangle   \ .
\end{eqnarray}
As a matter of fact, it has been shown in
\cite{MR2277640} that $\widehat{T}$ is the most general operator which
is tridiagonal in both bases in finite-dimensional situations.

Let  $\widehat{T}_{mn} = \langle m | \widehat T | n 
\rangle$, and  define the ``chopped'' matrix $T$ by
\begin{equation}
	T=|\widehat T_{mn}|_{0\leq m,n \leq \ell} \,.
	\label{Tmat}
\end{equation}	
Following the results of \cite{Perline:1987:DTL:37170.37175, 2018CMaPh.364.1041G}, we know that 
$T$ and $C$ will commute
\begin{equation}\label{eq:comTC}
 [T,C]=0
\end{equation}
if the parameters in $\widehat T$ \eqref{Heun} are given by
\begin{equation}\label{eq:cons1}
    \tau=0\ ,\quad  
	\mu= -(\omega_K+\omega_{K+1}) \quad \text{and} \qquad 
	\nu= -(\lambda_\ell+\lambda_{\ell+1}) \ .
\end{equation}
Indeed, with the particular value of $\nu$ given by \eqref{eq:cons1},
we see that the matrix $\widehat T$ leaves the subspace $\{|n\rangle,
n=0,1,\dots, \ell\}$ invariant.  Therefore 
$T$ commutes with $\pi_1$.
Similarly, with the particular value of $\mu$ given by
\eqref{eq:cons1}, $\widehat T$ leaves the subspace
$\{|\omega_k\rangle, k=0,1,\dots, K\}$ invariant.  Therefore $T$ commutes
with $\pi_2$.  Finally, in view of \eqref{Cpi},
it is easy to get the result \eqref{eq:comTC}.

The main result of this section is the tridiagonal matrix $T$ \eqref{Tmat} 
i.e.
\begin{eqnarray}
 T =\begin{pmatrix}
             d_{0} & t_{0} & \\
             t_{0} & d_{1} & t_{1} \\
             & t_{1} & d_{2} & t_{2} \\
             && \ddots & \ddots & \ddots \\
            &&&t_{\ell-2} & d_{\ell-1} & 
			t_{\ell-1}\\
             &&&& t_{\ell-1} & d_{\ell}
            \end{pmatrix}\,,
			\label{Tmatfinal}
\end{eqnarray}
which commutes with the correlation matrix \eqref{eq:comTC}
and whose nonzero matrix elements are given by (see \eqref{Tpos}) 
\begin{align}
t_{n} &= J_n(\lambda_n+\lambda_{n+1}-\lambda_\ell-\lambda_{\ell+1}) 
\,, \non\\
d_{n} &=- B_n (2\lambda_n - \lambda_\ell - \lambda_{\ell+1})
- \lambda_n (\omega_K+\omega_{K+1})\,.
\label{Tmatelems}
\end{align}	  
A key ingredient obviously is the operator
$\widehat X$ defined in \eqref{Xmat}.  In the following sections, we
apply this construction to both homogeneous and inhomogeneous 
free-Fermionic chains.

If $t_n\neq 0$ (which is the case in the examples below), 
$T$ is non-degenerate (see e.g. Lemma 3.1 in \cite{2004math......8390T}) and
the commuting matrices $T$
and $C$ have a unique set of common eigenvectors.  Since $T$ is
tridiagonal, its eigenvectors can be readily computed numerically.
By acting with $C$ on these eigenvectors, the eigenvalues of $C$ can
be easily obtained.  The eigenvalues of the entanglement Hamiltonian
${\cal H}$, and therefore the entanglement entropy of the model, can
then also be easily obtained.

\section{The homogeneous chain}\label{sec:homog}

Let us construct the tridiagonal matrix $T$ \eqref{Tmatfinal} 
for the homogeneous chain, for which
\begin{equation}	
J_{0}= \ldots = J_{N-1} = -\frac{1}{2} \,, \qquad B_{n} = 0 \,.
\label{JBCheb}
\end{equation}
We make use of a certain discretization of the (continuous) Chebyshev 
polynomials of the second kind, which are defined by 
(see e.g. \cite{Chebyshev, 2019arXiv190403789Z})
\begin{equation}
R_{n}(x) = \frac{\sin(\theta(n+1))}{\sin(\theta)} \,, \qquad x = 
\cos(\theta) \,, \qquad n = 0, 1, \ldots \,,
\end{equation}
which are polynomials in $x$ of degree $n$. Note that
$x$ is not restricted here to integer values.
These polynomials satisfy the recurrence relation (c.f. \eqref{recurgen})
\begin{equation}
2x R_{n}(x) = R_{n+1}(x) + R_{n-1}(x) \,, \qquad n = 0, 1, \ldots \,.
\label{recurCheb}
\end{equation}
Comparing the recurrence relations \eqref{recurphi} with $0 \le n \le 
N-1$ and \eqref{recurCheb}, and recalling the parameter values \eqref{JBCheb}, we 
see that $\phi_n(\omega_k) \propto R_{n}(x)$. Moreover, the recurrence 
relation \eqref{recurphi} with $n=N$ leads to the constraint
\begin{equation}
	2\cos(\theta) \sin((N+1)\theta) = \sin(N\theta) \,,
\end{equation}	
which has solutions
\begin{equation}
\theta = \theta_{k} = \frac{\pi (k+1)}{N+2}
\end{equation}
for any integer $k$. Imposing the normalization \eqref{orthophi}, we 
conclude that the eigenfunctions are given by
\begin{equation}	
\phi_n(\omega_k) = \sqrt{\frac{2}{N+2}} \sin (\theta_{k}) R_{n}(x_{k})
= \sqrt{\frac{2}{N+2}} \sin \left[\frac{\pi 
(k+1)(n+1)}{N+2}\right]\,, 
\label{phiCheb}
\end{equation}
where
\begin{equation}
\omega_k = -x_{k} = -\cos(\theta_{k}) \,, \qquad k=0, 1, \ldots, N \,.
\end{equation}

Starting from the recurrence relation for $\phi_n(\omega_{k})$, we 
can relabel $n \leftrightarrow k$ and use the property $\phi_n(\omega_{k}) = 
\phi_k(\omega_{n})$ of the eigenfunctions \eqref{phiCheb}
to obtain the difference relation 
\begin{equation}
	\omega_{n} \phi_n(\omega_{k}) = -\frac{1}{2}\phi_n(\omega_{k+1}) 
	-\frac{1}{2}\phi_n(\omega_{k-1}) \,,
\end{equation}	
c.f. \eqref{diffphi}. We can therefore define $\widehat X$ as in \eqref{Xmat}, with
\begin{equation}
	\lambda_{n} = \omega_{n} = -\cos(\theta_{n}) \,.
\end{equation}	
The matrix $T$ is therefore given by \eqref{Tmatfinal}, with 
\begin{align}
	t_{n} &= \frac{1}{2}\left[\cos(\theta_{n})+\cos(\theta_{n+1}) - 
	\cos(\theta_{\ell}) - \cos(\theta_{\ell+1})\right]\,, \non \\
	\qquad d_{n} &= -\cos(\theta_{n}) \left[ \cos(\theta_{K})+ 
	\cos(\theta_{K+1})\right]\,.
\end{align}
This result agrees (up to 
overall and additive constants, accounting for differences in 
conventions) with the recent
result for the same model in \cite{2018arXiv180500078E}
(see also \cite {Cardy:2016fqc}). Our new observation is that 
these results follow from the application of the algebraic Heun construction to truncated Chebychev
polynomials of the second kind.

\section{Inhomogeneous chains}\label{sec:inhomog}

We now turn to some examples of inhomogeneous chains. We consider 
models corresponding to Krawtchouk and dual Hahn polynomials in  
Secs. \ref{subsec:Krawtchouk} and \ref{subsec:dualHahn}, respectively.
Let us mention that the commuting matrices associated to these polynomials were first obtained by Perlstadt \cite{doi:10.1137/0604012,doi:10.1137/0515081}
and recovered algebraically by Perline \cite{Perline:1987:DTL:37170.37175}.

\subsection{Krawtchouk}\label{subsec:Krawtchouk}

The Krawtchouk polynomials, which in general depend on one parameter ($p$), are defined by \cite{Koekoek,Koekoek2010}
\begin{equation}
	\setlength\arraycolsep{1pt}
	R_{n}(\lambda(x)) = {}_{2} F_{1}\left(\begin{matrix} -n, & & -x
	\\&-N  &\end{matrix}\ 
	;\frac{1}{p}\right)\,, \qquad n = 0, 1, \ldots, N\,,
	\label{Krawtchouk}
\end{equation}	
where 
\begin{equation}
	\lambda(x) = -x \,.
\end{equation}	
The orthogonality relation is given by \eqref{orthoW} with \footnote{The 
Pochhammer (or shifted factorial) symbol $(a)_{k}$ is defined by
\begin{equation}
	(a)_{0} = 1\,, \qquad (a)_{k} = a (a+1)(a+2)  \cdots (a+k-1) \,, 
	\quad k = 1, 2, \ldots \,. \non
\end{equation}
We note the identity
\begin{equation}
	\frac{(-N)_{n} (-1)^{n}}{n!} = {N\choose n} \,. \non 
\end{equation}
	}
\begin{equation}
	W(x) = {N \choose x} p^{x} (1-p)^{N-x}\,, \qquad
	U_{n} = \left(\frac{1-p}{p}\right)^{n}/{N\choose n}\,,
\end{equation}	
for $0<p<1$. The recurrence relation is given by \eqref{recurgen} with
\begin{equation}
	A_{n} = p(N-n) \,, \qquad
	C_{n} = n(1-p) \,,
	\label{ACK}
\end{equation}	
while the difference relation is given by \eqref{diffgen} with 
\begin{equation}
\overline{A}(x) = p (N-x)	\,, \qquad
\overline{C}(x) = x (1-p) \,, \qquad f(n) = -n \,.
\label{barAbarCK}
\end{equation}	
Note that the Krawtchouk polynomials \eqref{Krawtchouk} are 
self-dual: they are invariant under the interchange $n 
\leftrightarrow x$. Hence, the coefficients \eqref{ACK} and 
\eqref{barAbarCK} are related by 
$A \leftrightarrow \overline{A}$ and $C \leftrightarrow \overline{C}$ 
under this interchange.

The parameters in the corresponding Hamiltonian are given by 
\eqref{JBparams} \footnote{We choose $\varepsilon=1$, 
and we introduce in $B_{n}$ an extra factor $-1$ in 
order to ensure  $\omega_{k} < \omega_{k+1}$.}
\begin{equation} 
	J_{n} =  \sqrt{(N-n)(n+1)p(1-p)}\,, \qquad
	B_{n} = -\left[N p + n(1 - 2 p)\right]\,,
\end{equation}
which corresponds to an inhomogeneous chain.
For simplicity, we henceforth consider the special case $ p = 
\frac{1}{2}$, for which the chain is mirror symmetric and admits end-to-end perfect state transfer 
\cite{2004PhRvL..93w0502A, 2009arXiv0903.4274K,2010JPhA...43h5302C,2012PhRvA..85a2323V}.
The $\alpha$'s are then given by \eqref{alphas}
\begin{equation}
	\alpha_{n} = \alpha_{0}/\sqrt{{N\choose n}} 
	= 1 /\sqrt{{N\choose n}} \,,
\end{equation}
where $\alpha_{0}=1$ has been chosen to ensure the normalization in 
\eqref{orthophi}. The eigenfunctions $\phi_n(\omega_k)$ are given by
\eqref{Rphi}
\begin{equation}
\phi_n(\omega_k) =  (-1)^{n} 
2^{-\frac{N}{2}}\sqrt{{N\choose n} {N\choose k}}\,
R_{n}(\lambda(k)) \,,
\end{equation}
where
\begin{equation}
	\omega_{k} = -\lambda(k) = k \,.
\end{equation}	
The difference relation is given by \eqref{diffphi}, with
\begin{equation}
\overline{J}_{k} = 
-\frac{1}{2}\sqrt{(N-k)(k+1)}\,, \qquad \overline{B}_{k} = 
-\frac{N}{2}\,, \qquad \lambda_{n} = n \,.
\end{equation} 
The matrix $T$ is therefore of the form \eqref{Tmatfinal}, with 
\begin{align}
	t_{n} &= (n-\ell)\sqrt{(N-n)(n+1)}\,, \non \\
	d_{n} &= \frac{N}{2}(2n-2\ell-1) - n(2K+1) \,.
\end{align}

\subsection{Dual Hahn}\label{subsec:dualHahn}

The dual Hahn polynomials, which in general depend on two parameters ($\gamma\,, 
\delta$), are defined by \cite{Koekoek,Koekoek2010}
\begin{equation}
	\setlength\arraycolsep{1pt}
	R_{n}(\lambda(x)) = {}_{3} F_{2}\left(\begin{matrix} -n, & & -x, 
	& & x+ \gamma + \delta +1 \\&\gamma+1, & & -N &\end{matrix}\ 
	;1\right)\,, \qquad n = 0, 1, \ldots, N\,,
\end{equation}	
where 
\begin{equation}
	\lambda(x) = x (x + \gamma + \delta + 1) \,.
\end{equation}	
 They obey the orthogonality relation \eqref{orthoW} with 
\begin{equation}
	W(x) = \frac{(2x+\gamma+\delta+1)(\gamma+1)_{x} N!}
	{(x+\gamma+\delta+1)_{N+1} (\delta+1)_{x}}{N \choose x}\,, \qquad
	U_{n} = \left[{\gamma+n\choose n}{\delta+N-n\choose N-n} 
	\right]^{-1}\,,
	\end{equation}	
for $\gamma\,, \delta >-1$ or $\gamma\,, \delta < -N$. 
These polynomials satisfy the recurrence relation \eqref{recurgen} with
\begin{equation}
	A_{n} = (n + \gamma +1)(n - N) \,, \qquad
	C_{n} = n(n - \delta - N-1) \,,
\end{equation}	
and the difference relation \eqref{diffgen} with $f(n) 
= -n$ and 
\begin{equation}
\overline{A}(x) = \frac{(x + \gamma + 1)(x + \gamma + \delta + 1)(N-x)}	
{(2x + \gamma + \delta + 1)(2x + \gamma + \delta + 2)}	\,, \qquad
\overline{C}(x) = \frac{x(x + \gamma + \delta + N + 1)(x + \delta )}	
{(2x + \gamma + \delta )(2x + \gamma + \delta + 1)} \,.
\end{equation}	

The parameters in the corresponding Hamiltonian read (choosing 
$\varepsilon=-1$) by \eqref{JBparams}
\begin{equation}
	J_{n} = - \sqrt{(n+1)(n + \gamma +1)(N-n)(N + \delta -n)} \,, \qquad
	B_{n} = - N - (N-n)(2n + \gamma) - n \delta \,,
\end{equation}
which also corresponds to an inhomogeneous chain.
For simplicity, we henceforth consider the special case $ \delta = 
\gamma >0$. The $\alpha$'s of \eqref{alphas} are then
\begin{equation}
	\alpha_{n} = \alpha_{0} \sqrt{\frac{n! 
	{N+\gamma\choose n}}{{N\choose n} (\gamma+1)_{n}}} 
	= \sqrt{\frac{N! n! 
	{N+\gamma\choose n}}{{N\choose n}(\gamma+1)_{N}(\gamma+1)_{n}}}\,,
\end{equation}
where $\alpha_{0}$ has been chosen to ensure the normalization in 
\eqref{orthophi} for the eigenfunctions $\phi_n(\omega_k)$, which obey
the recursion relation \eqref{recurphi}, and which are given by
\eqref{Rphi}
\begin{equation}
\phi_n(\omega_k) =  \left[
\frac{{N\choose n}{N\choose k}(2k + 2\gamma+1)(\gamma+1)_{N} 
(\gamma+1)_{n}}{n! {N+\gamma\choose n}(k + 
2\gamma+1)_{N+1}}\right]^{1/2}
R_{n}(\lambda(k)) \,,
\end{equation}
where
\begin{equation}
	\omega_{k} = \lambda(k) = k (k + 2\gamma+1) \,.
\end{equation}	
These eigenfunctions obey the difference relation \eqref{diffphi}, with
\begin{equation}
\overline{J}_{k} = 
\frac{1}{2}\sqrt{\frac{(N-k)(k+1)(k+2\gamma+1)(N+k+2\gamma+2)}
{(2k+2\gamma+1)(2k+2\gamma+3)}}\,, \qquad \overline{B}_{k} = 
\frac{N}{2}\,, \qquad \lambda_{n} = -n \,.
\end{equation} 
The matrix $T$ is therefore of the form \eqref{Tmatfinal}, with 
\begin{align}
	t_{n} &= -2(\ell-n)\sqrt{(n+1)(n + \gamma +1)(N-n)(N + \gamma -n)} 
	\,, \non \\
	\qquad d_{n} &= (2\ell-2n+1)\left[N(\gamma+1)+2N n -2 n^{2}\right]
	+ 2 n \left[ \gamma+1 +K(K+2\gamma+2)\right]\,.
\end{align}

\section{Conclusions}\label{sec:conclude}

For any free-Fermion chain associated with a discrete orthogonal
polynomial, we have constructed a tridiagonal matrix $T$ that
commutes with the ``chopped'' correlation matrix $C$, and hence, with 
the hopping matrix for the 
entanglement Hamiltonian. This matrix $T$ is nothing but a specialization of the 
algebraic Heun operator. The provenance of this construction is the 
remarkable fact that the wavefunctions (orthogonal polynomials) obey both 
recurrence and difference relations with three terms. We expect that 
this result will facilitate the computation of the finite-size 
entanglement entropy for such models.

We ``chopped'' here in frequency by keeping only the momentum modes in the 
interval $[0\,, K]$, see e.g. \eqref{Cdag}. It would be interesting to 
know whether such a matrix $T$ can still be constructed if one chops 
in other ways, such as in an arbitrary interval $[K_{1}\,, K_{2}]$, or 
in more than one disjoint intervals, etc.

Free-Fermion chains are simple examples of quantum integrable models. It 
would be instructive to explore whether similar constructions are 
possible for interacting quantum integrable models. 
An attractive candidate would be the XXZ spin 
	chain with $\Delta=\pm \frac{1}{2}$, see e.g. \cite{Colomo:2004jw}.

\paragraph{Acknowledgements} 
We much benefitted from discussions with A. Gonzalez-Lopez, W. 
Witczak-Krempa and A. Zhedanov; 
and we thank A. Gr\"unbaum for correspondence and especially for bringing \cite {2018arXiv180500078E} to our attention.
N. Cramp\'e is gratefully holding a CRM--Simons professorship.  R.
Nepomechie warmly thanks the Centre de Recherches Math\'ematiques
(CRM) for hospitality and support during his visit to Montreal in the
course of this investigation.  The research of L. Vinet is supported
in part by a Discovery Grant from the Natural Science and Engineering
Research Council (NSERC) of Canada.


\providecommand{\href}[2]{#2}\begingroup\raggedright\endgroup

\end{document}